\documentclass[journal,comsoc]{IEEEtran}

\usepackage[T1]{fontenc} 

\ifCLASSINFOpdf
\else
\fi
\usepackage{amsmath}

\usepackage[cmintegrals]{newtxmath}

\usepackage{amsfonts}
\usepackage{bm}
\usepackage[fleqn,tbtags]{mathtools}
\usepackage{algorithm}
\usepackage{algpseudocode}
\usepackage{url}
\usepackage{graphicx}
\usepackage[caption=false,font=footnotesize]{subfig}

\usepackage[dvipdfm,colorlinks,linkcolor=black,anchorcolor=black,citecolor=black]{hyperref}

\newcommand{\vect}[1]{\mathbf{#1}}
\newcommand{\mat}[1]{\mathbf{#1}}
\newcommand{\field}[1]{\mathbb{#1}}
\newcommand{\C}{\field{C}}

\newcommand{\abs}[1]{\left\lvert#1\right\rvert}
\newcommand{\norm}[1]{\left\lVert#1\right\rVert}

\newcommand{\CT}{\text{\textit{H}}}
\newcommand{\st}{\quad \text{s.t.} \quad }

\DeclareMathOperator{\grad}{grad}
\DeclareMathOperator{\Real}{Re}
\DeclareMathOperator{\ddiag}{ddiag}
\DeclareMathOperator{\Tr}{Tr}

\DeclareMathOperator{\SINR}{\rho}

\begin{document}
%
\title{Coordinated Multicell Multicast Beamforming Based on Manifold Optimization}
%
%
%

\author{Longfei~Zhou,~\IEEEmembership{Student Member,~IEEE,}
	 Le~Zheng,~\IEEEmembership{Member,~IEEE,}
	Xiaodong~Wang,~\IEEEmembership{Fellow,~IEEE,} \\
	Wei~Jiang,~\IEEEmembership{Member,~IEEE,}
	and~Wu~Luo,~\IEEEmembership{Member,~IEEE}
	
	\thanks{This work was supported
		by the National Natural Science Foundation of China under grant 61171080. } 
	\thanks{L. Zhou, W. Jiang, and W. Luo are with the State Key Laboratory of Advanced
		Optical Communication Systems and Networks, Peking University,Beijing 100871, China 
		(email: \{zhoulongfei, jiangwei, luow\}@pku.edu.cn).}
	\thanks{L. Zheng and X. Wang are with the Department of Electrical Engineering, Columbia University, 
		New York, NY 10027 USA (e-mail: le.zheng.cn@gmail.com; wangx@ee.columbia.edu).} 
}

\maketitle

\begin{abstract}
Multicast beamforming is a key technology for next-generation wireless cellular networks
to support high-rate content distribution services. 
In this paper, the coordinated downlink multicast beamforming design in
 multicell networks is considered.
The goal is to maximize the minimum signal-to-interference-plus-noise ratio of all users  
under individual base station power constraints.
We exploit the fractional form of the objective function and geometric properties 
of the constraints to reformulate the problem as a parametric manifold optimization program.
Afterwards we propose a low-complexity Dinkelbach-type algorithm combined with adaptive exponential smoothing
and Riemannian conjugate gradient iteration, which is guaranteed to converge.
Numerical experiments show that the proposed algorithm outperforms the existing SDP-based method
and DC-programming-based method and achieves near-optimal performance. 
\end{abstract}


%

\begin{IEEEkeywords}
Multicast beamforming, max-min fair, manifold optimization, Riemannian conjugate gradient.
\end{IEEEkeywords}

\section{Introduction}
\IEEEPARstart{E}{xplosive} demands for high-rate wireless content distribution services,
such as audio and video streaming, software updates, and Internet TV, have  
motivated extensive research on advanced physical layer techniques 
to boost the capacity of wireless networks \cite{kong2016embracing}.
Downlink multicast beamforming is a powerful technique to improve the wireless throughput
for next generation cellular networks. %

A variety of multicast beamforming problems have been investigated for different scenarios.
For single-cell system, single-group multicast beamforming was first discussed in \cite{2006TSP-SDR},
where all users request a common information from the base station (BS), 
and then extended to multi-group multicast in \cite{2008TSP-MultiGroup}.
Recently, the multi-group multicast beamforming under per-antenna power constraints was further
investigated in \cite{2014-multicastPAPC}.
Moreover, the coordinated multicast beamforming with individual BS power constraints in
multi-cell networks has been considered in \cite{2013-coordinated}.
Some other issues, such as energy efficient design and user selection, were also studied in
\cite{2015-EnergyEfficientMulticast} \cite{2015-coordinatedMulticastingUserSelection} \cite{hsu2016joint}.

In this paper, we revisit the max-min fair coordinated multicast beamforming problem in \cite{2013-coordinated}.
The existing approach of semidefinite relaxation (SDR) and Gaussian randomization in 
\cite{2008TSP-MultiGroup}\cite{2013-coordinated} has following drawbacks. 
First, SDR is not scalable to large-scale antenna systems 
as the number of involved variables is quadratic in the number of antennas.
Second, extracting a rank-one component from the optimum solution to the SDR problem is NP-hard in general.
The polynomial-time approximation method of Gaussian randomization in 
\cite{2008TSP-MultiGroup}\cite{2013-coordinated}
needs to solve a large number of multicast feasibility power control subproblems
and such approximation degrades considerably as the number of antennas increases \cite{2006TSP-SDR}.
In this paper, we present a new max-min fair multicast beamforming design that outperforms the existing 
methods in \cite{2013-coordinated} and \cite{hsu2016joint} yet with a much lower complexity.

\section{System Model and Problem Formulation}
Consider a multicell multicast scenario consisting of $L$ cells and $K$ single-antenna users per cell,
sharing a common time-frequency resource.
Each cell has a BS equipped with $M$ antennas.
The BS in the $l$-th cell uses an $M\times1$ beamforming vector $\vect{\tilde{w}}_l$ 
to send a zero-mean and unit-variance multicast signal $s_l$ to all users in the $l$-th cell.
The signal received by the $k$-{th} user in the $l$-th cell is 
\begin{equation}\label{rcvdsignal}
y_{l,k} = \vect{\tilde{h}}_{l,l,k}^\CT\vect{\tilde{w}}_l s_l + 
\sum_{j=1,j\ne l}^{L}\vect{\tilde{h}}_{j,l,k}^\CT\vect{\tilde{w}}_j s_j +n_{l,k},
\end{equation}
where $\vect{\tilde{h}}_{j,l,k} \in \C^{M\times1}$ is the channel between the 
$k$-{th} user in the $l$-th cell and the BS in the $j$-th cell.
$n_{l,k}\sim {\cal CN}(0,\sigma_{l,k}^2)$ is the additive white Gaussian noise (AWGN) 
at the $k$-{th} user in the $l$-th cell and is independent of $\vect{\tilde{h}}_{j,l,k}$ and $s_l$.

Assume that a central processing unit collects the channel state information between all BSs
and all users in the system. Based on the received signal model in \eqref{rcvdsignal}, the performance
of each user can be characterized by the signal-to-interference-plus-noise ratio (SINR). 
The problem of interest is to maximize the minimum weighted SINR
among all users under individual BS power constraints
\begin{subequations}\label{MMNF}
\begin{align}
(\mathcal{F}):\qquad  \max_{\mat{\tilde{W}} \in \C^{M\times L}} \min_{l,k} &\quad  
\frac{1}{\Gamma_l}
\frac{\abs{\vect{\tilde{h}}_{l,l,k}^\CT\vect{\tilde{w}}_l}^2}
{\sum_{j\ne l}^{L}\abs{\vect{\tilde{h}}_{j,l,k}^\CT\vect{\tilde{w}}_j}^2+\sigma_{l,k}^2} \\
\st & \norm{\vect{\tilde{w}}_l}_2^2 \le P_l \:\forall l, \label{powerConstrnt}
\end{align}
\end{subequations}
where $\mat{\tilde{W}} = [\vect{\tilde{w}}_1,\dots,\vect{\tilde{w}}_L] \in \C^{M\times L},$
$\vect{\Gamma}=[\Gamma_1,\Gamma_2,\dots,\Gamma_L]^T$ with each entry 
$\Gamma_l$ being the target SINR for all users in the $l$-th cell, and
$\vect{P}=[P_1,\dots,P_L]^T$ is the power budget vector for all BSs. 
Since the multicast information
rate for users within one cell is the same, we set a common target SINR value for all 
users in the same cell.

\section{Algorithm Design}
\subsection{Preliminary Analysis}
We first map the feasible region \eqref{powerConstrnt} onto spheres by introducing $L$ complex slack variables 
$\tilde{w}_{l,M+1},l=1,\dots,L,$ such that $\abs{\tilde{w}_{l,M+1}}^2+\norm{\vect{\tilde{w}}_l}_2^2=P_l, \forall l.$
Let $\vect{w}_l=\tfrac{1}{\sqrt{P_l}} [\vect{\tilde{w}}_l^T,\tilde{w}_{l,M+1}]^T \in \C^{(M+1)\times1}$
and $\mat{W} = [\vect{w}_1,\dots,\vect{w}_L] \in \C^{(M+1)\times L},$
then the feasible region \eqref{powerConstrnt} becomes 
$\mathsf{S}=\{\mat{W}\in\C^{(M+1)\times L}|\norm{\vect{w}_l}_2^2 = 1, \forall l\}.$
Next, channel vector $\vect{\tilde{h}}_{j,l,k}$ is normalized by $P_j$ and $\sigma_{l,k}$ as follows, 
$\vect{h}_{j,l,k} = \tfrac{\sqrt{P_j}}{\sigma_{l,k}} [\vect{\tilde{h}}_{j,l,k}^T,0]^T \in \C^{(M+1)\times1},\forall j,l,k.$ 
Denote
\begin{align}\label{def_of_rho}
\SINR(\mat{W}) = \min_{l,k} \frac{1}{\Gamma_l}\frac{\abs{\vect{h}_{l,l,k}^\CT\vect{w}_l}^2 }
{\sum_{j\ne l}^{L}\abs{\vect{h}_{j,l,k}^\CT\vect{w}_j}^2+1}.
\end{align}
The multicast beamforming problem $(\mathcal{F})$ is rewritten as
\begin{align}
(\mathcal{F}_*):\qquad\max_{\mat{W}} \: \SINR(\mat{W}) \st \mat{W} \in \mathsf{S}.
\end{align}
Denote $F(\mat{W},t) = \min\limits_{l,k} f_{l,k}(\mat{W},t),$ where 
\begin{align}\label{def_of_f}
f_{l,k}(\mat{W},t) = \frac{1}{\Gamma_l}\abs{\vect{h}_{l,l,k}^\CT\vect{w}_l}^2 
-t\left(\sum_{j\ne l}^{L}\abs{\vect{h}_{j,l,k}^\CT\vect{w}_j}^2+1\right)
\end{align}
represents the received power shortage or redundancy of 
the $k$-{th} user in the $l$-th cell to achieve a weighted SINR value of $t$.  
We reformulate the problem $(\mathcal{F}_*)$  as 
the following parametric programming problem
\begin{align}\label{NSO}
(\mathcal{F}_t):\qquad F(t)=\max_{\mat{W}} \: F(\mat{W},t) \st \mat{W} \in \mathsf{S}.
\end{align}

Note $\SINR(\mat{W})$ has a fractional form, while $F(\mat{W},t)$ is the point-wise minimum of quadratic functions. 
In general, $F(\mat{W},t)$ is easier to handle than  $\SINR(\mat{W}).$
Maximizing $\SINR(\mat{W})$ over $\mathsf{S}$ is equivalent to maximizing 
$F(\mat{W},t)$ over $\mathsf{S},$ i.e., calculating the single-variable function $F(t)$ defined in \eqref{NSO},
since $F(t)$  is continuous and strictly decreasing over $[0,\infty),$
and the unique zero of $F(t)$ is the optimal objective value of $(\mathcal{F}_*).$	
Thus solving $(\mathcal{F}_*)$ reduces to determining the unique value of $t$ such that $F(t)=0$,
which can be achieved by solving a sequence of parametric subproblems $(\mathcal{F}_{t})$ for different values of $t.$

It is desirable that the sequence of parameters ${t_k}$ increases monotonically, since this guarantees
that each solution of $(\mathcal{F}_{t_k})$ is a feasible point and an appropriate initial 
point as well for the next subproblem.
Bisection methods do not have this property, and therefore we  
employ the Dinkelbach-type  algorithm to find the root of $F(t)$.
The basic idea is to first find $t_k$ such 
that $F(\mat{W}^{(k)},t_k)=0$ for a given $\mat{W}^{(k)} \in \mathsf{S}$ and 
then find a solution $\mat{W}^{(k+1)}$ to $(\mathcal{F}_{t_k})$ \cite{1991-crouzeix-GFP}. 

Computing $F(t)$ or solving $(\mathcal{F}_{t})$ is central to solving the original problem $(\mathcal{F}_*)$.
However, $(\mathcal{F}_t)$ for a given $t$ is equivalent to a 
non-convex quadratically constrained quadratic program (QCQP), which is difficult to solve \cite{2008TSP-MultiGroup}.
Instead, we propose to compute a stationary solution of $(\mathcal{F}_{t})$ 
that provides an achievable low bound of $F(t)$. 
Since $F(\mat{W},t)$ is non-smooth, we turn to the following smoothed surrogate problem
\begin{align}\label{SO}
(\mathcal{F}_{t,\mu}):\qquad\max_{\mat{W}} \: F(\mat{W},t,\mu) \st \mat{W} \in \mathsf{S},
\end{align}
where $\mu>0$ is a smoothing parameter and  
\begin{align}\label{SmoothingObj}
F(\mat{W},t,\mu) = -\mu\log\sum_{l,k}\exp{(-f_{l,k}(\mat{W},t)/\mu)}
\end{align}
is the exponential smoothing of $F(\mat{W},t)$ that satisfies \cite{2003-polak-algorithms}
\begin{align}\label{BoundInequality}
F(\mat{W},t,\mu) \le F(\mat{W},t) \le  F(\mat{W},t,\mu) + \mu \log (KL).
\end{align}
Moreover, $F(\mat{W},t,\mu)$ increases 
while $F(\mat{W},t,\mu) + \mu \log (KL)$ decreases, as $\mu$ decreases.
Therefore, a small $\mu$ leads to high approximation accuracy. 
However, when $\mu$ is small, 
the problem $(\mathcal{F}_{t,\mu})$ is nearly ill-conditioned, 
which is difficult to solve.
An effective strategy is to solve a 
sequence of gradually more accurate approximations \cite{2003-polak-algorithms}.
Taking into consideration the structure of the constraint set $\mathsf{S},$
a Riemannian conjugate gradient (RCG) method is well suitable for obtaining a stationary solution 
to $(\mathcal{F}_{t,\mu})$ with low complexity, which will be detailed in the next subsection.

As mentioned above, we use a Dinkelbach-type procedure to solve $(\mathcal{F}_*)$ as follows: 
given $\mat{W}^{(k)} \in \mathsf{S}$, set  $t_k = \SINR(\mat{W}^{(k)})$ and 
then find a solution $\mat{W}^{(k+1)} \in \mathsf{S}$ to 
 $(\mathcal{F}_{t_k,\mu})$ for some $\mu$ such that 
$ F(\mat{W}^{(k+1)},t_k) > F(\mat{W}^{(k)},t_k).$
The latter is always possible by the RCG method for a sufficiently small $\mu$ 
as long as $\mat{W}^{(k)}$ is not a stationary point of $(\mathcal{F}_{t_k}),$ 
due to \eqref{BoundInequality}.
To obtain such a solution, the RCG method would be repeatedly applied on the
problems $(\mathcal{F}_{{t_k},\mu})$ for a decreasing sequence of $\mu$ values.
An important feature of the proposed Dinkelbach-type procedure is the monotonicity of the generated 
sequence $\{t_k\}_{k=0}^{+\infty},$ which guarantees convergence and makes the algorithm 
numerically more stable. 
For clarity, the proposed Dinkelbach-type procedure that employs the RCG method 
(line 1) is summarized in Algorithm 1. 
 \begin{algorithm}
	\caption{DT-RCG algorithm for problem $(\mathcal{F}_*)$}
	\textbf{input}: initial point $\mat{W}^{(0)} \in \mathsf{S}, \mu_0$ \\
	\textbf{output}: $\mat{W}^{(k)} $ and $t_k$
	\begin{algorithmic}
		\State \textbf{Initialization}   Set $t_0 = \SINR(\mat{W}^{(0)}).$
		\For{$k=1,2,\dots$}
		\State \textbf{1.} $\mat{W}^{(k)}=\text{RiemannianConGrad}(\mat{W}^{(k-1)},t_{k-1},\mu_{k-1}).$
		\State \textbf{2.} \textbf{If} $F(\mat{W}^{(k)},t_{k-1}) > F(\mat{W}^{(k-1)},t_{k-1})$ \textbf{then} \\
		\qquad\qquad $t_{k}=\SINR(\mat{W}^{(k)}),\mu_{k} = \mu_{k-1};$ \\
		\qquad\: \textbf{else}\\ \qquad\qquad $\mat{W}^{(k)} = \mat{W}^{(k-1)},t_{k}=t_{k-1},\mu_{k} = \mu_{k-1}/2.$
		\State \textbf{3.} \textbf{If} $\mu_{k} < \varepsilon$ \textbf{then} STOP.
		\EndFor
	\end{algorithmic} 
\end{algorithm} 
\subsection{Riemannian Conjugate Gradient Algorithm}
Motivated by the superior performance of nonlinear conjugate gradient methods to
large-scale unconstrained optimization problems \cite{2006-hager-NCGsurvey},
we treat the surrogate problem $(\mathcal{F}_{t,\mu})$ as an unconstrained optimization problem 
on a complex oblique manifold and devise a RCG algorithm to 
find an approximate stationary solution by using the framework of retraction-based manifold optimization \cite{boumal2015low}\cite{2009ManOpt}.
To simplify notation, $F(\mat{W},t,\mu)$ will be simply denoted by $F(\mat{W})$ in this subsection.

Conceptually, the RCG algorithm has three stages in each iteration: (\romannumeral1) Compute the Riemannian gradient, i.e.,the tangent vector in the tangent space corresponding to the direction of steepest ascent of $F(\mat{W});$ 
(\romannumeral2) Find a tangent vector that is conjugate to the Riemannian gradient as the search direction; 
(\romannumeral3) Invoke the metric projection as a retraction that maps a tangent vector to a point on the manifold. 
We will next detail some key ingredients of the RCG algorithm.
Background on manifold optimization can be found in \cite{2009ManOpt}.

A \textit{manifold} $\mathcal{M}$ is a topological space that
resembles a Euclidean space near each point. 
For our problem, the feasible set $\mathsf{S}$ defines a \textit{complex oblique manifold},
namely, the Cartesian product of unit spheres 
\begin{equation}
\mathcal{M} =\left\{\mat{W}\in \C^{(M+1)\times L}|\ddiag(\mat{W^\CT W})=\mat{I}_L\right\},
\end{equation}
where $\ddiag(\mat{Z})$ forms a diagonal matrix, whose diagonal elements are those of $\mat{Z}$.
The \textit{tangent vector} of any smooth
curve through the point $\mat{W}$ characterizes the direction along which it can move. 
All tangent vectors at a given point on manifold form a linear subspace, called \textit{tangent space}.
In our case, the tangent space $T_\mat{W}\mathcal{M}$ at the point $\mat{W}\in \mathcal{M}$ is described by
\begin{equation}
T_\mat{W}\mathcal{M} = \left\{\mat{U}\in \C^{(M+1)\times L}| \ddiag(\Real(\mat{W}^\CT\mat{U}))= \mat{0} \right\}.
\end{equation}

To measure distances and angles on tangent space and use calculus on manifold, 
the canonical inner product $\langle \mat{U},\mat{V}\rangle_\mat{W} = \Real\left\{\Tr(\mat{U}^\CT\mat{V})\right\}$
is chosen as the \textit{Riemannian metric} on the tangent space $T_\mat{W}\mathcal{M},$
which makes $\mathcal{M}$ a \textit{Riemannian manifold}.
Hence, the \textit{Riemannian gradient} of $F(\mat{W})$ on $\mathcal{M},$ 
which is the unique tangent vector in the tangent space $T_\mat{W}\mathcal{M}$ 
that gives the largest increase in $F(\mat{W}),$ 
is given by the orthogonal projection of the Euclidean 
gradient $\nabla_\mat{W} F(\mat{W})$ onto  $T_\mat{W}\mathcal{M}$, i.e.,
\begin{equation}\label{grad_F}
\grad F(\mat{W}) = \nabla_\mat{W} F(\mat{W}) - \mat{W} \ddiag(\Real(\mat{W}^\CT\nabla_\mat{W} F(\mat{W}))).
\end{equation}
The Euclidean gradient $\nabla_\mat{W} F(\mat{W})$ is expressed as 
\begin{equation}
\nabla_\mat{W} F(\mat{W}) = \left[\frac{\partial F(\mat{W})}{\partial \vect{w}_1},\cdots,
\frac{\partial F(\mat{W})}{\partial \vect{w}_L} \right],
\end{equation}
where the complex-valued partial derivative $\frac{\partial F(\mat{W})}{\partial \vect{w}_l}$ is computed as
\begin{equation}
\frac{\partial F(\mat{W})}{\partial \vect{w}_l} = 2\sum_{m,k}
 a_{l,m}\beta_{m,k}(\mat{W})\vect{h}_{l,m,k}\vect{h}_{l,m,k}^\CT \vect{w}_l,
\end{equation}
with $a_{l,m} = -t$ for $l\ne m$ and $a_{l,m} = 1/\Gamma_l$ for $l=m,$ 
and $\beta_{m,k}(\mat{W})=\tfrac{ e^{-f_{m,k}(\mat{W},t)/\mu }}{\sum_{j,i}e^{-f_{j,i}(\mat{W},t)/\mu}}.$

The conjugate search direction is the weighted sum of the Riemannian gradient 
at present iteration and the search direction used at the previous iteration. 
However, two vectors in different tangent spaces cannot be added directly. 
This is accomplished by introducing the following \textit{vector transport} to map a 
tangent vector $\mat{U}\in T_{\mat{W}}\mathcal{M}$ to $T_{\mat{W}_{+}}\mathcal{M},$ 
\begin{equation}\label{vector_tanspt}
\mathcal{T}_{T_{\mat{W}_{+}}\mathcal{M}}(\mat{U}) =
\mat{U}-\mat{W}_{+}\ddiag(\Real(\mat{W}_{+}^\CT\mat{U})). 
\end{equation}
Specifically, given the previous and current Riemannian gradients, $\mat{G}_{*}$ and $\mat{G},$
and the previous conjugate direction $\mat{D}_{*},$ the current conjugate direction $\mat{D}$ is given by
\begin{equation}\label{conjugateDirection}
\mat{D}=\mat{G}+ \nu\mat{Y},
\end{equation}
where $\mat{Y}=\mathcal{T}_{T_{\mat{W}}\mathcal{M}}(\mat{D}_{*}),$ 
$\mat{Z}=\mathcal{T}_{T_{\mat{W}}\mathcal{M}}(\mat{G}_{*})$ and 
\begin{equation}\label{HSrule}
\nu = \max\left(0,
\frac{\langle \mat{G}_{}-\mat{Z},\mat{G}_{}\rangle_{\mat{W}}}
{\langle \mat{G}_{}-\mat{Z},\mat{Y}\rangle_{\mat{W}}}\right)
\end{equation} 
is the combination coefficient according to the modified Hestenes-Stiefel rule \cite{2006-hager-NCGsurvey}.

Given the conjugate direction, a so-called \textit{retraction}  mapping is used to 
map an element from the tangent space back to the
manifold. Retractions are essentially first-order approximations of the exponential map of the manifold. For our problem, the  following retraction, indeed the metric projection, is chosen to map a tangent 
vector $\mat{U}\in T_\mat{W}\mathcal{M}$ to  $\mathcal{M}$, 
\begin{equation}\label{retraction}
R_\mat{W}(\mat{U}) = \left(\mat{W+U}\right)\left(\ddiag\left(\left(\mat{W+U}\right)^\CT\left(\mat{W+U}\right)\right)\right)^{-1/2}.
\end{equation}

For completeness, the proposed RCG algorithm is summarized in Algorithm 2. 
To avoid confusion with the sequence generated by Algorithm 1,
the sequence in Algorithm 2 is denoted by $\{\mat{X}_n\}$. 
To guarantee the objective function to be non-decreasing in each iteration,
the RCG algorithm utilizes the Armijo line search in line 4 \cite{boumal2015low}. 
According to the convergence results for line-search method in \cite[Theorem 4.3.1]{2009ManOpt}, the RCG algorithm is 
guaranteed to globally converge to a stationary point of the surrogate problem $(\mathcal{F}_{t,\mu})$, 
namely, the point where the smoothed objective function $F(\mat{W},t,\mu)$ has vanishing Riemannian gradient.

\begin{algorithm}
	\caption{ $\mat{X}_{n} =\text{RiemannianConGrad}(\mat{X}_{0},t,\mu)$}
	\textbf{input}: initial point $\mat{X}_{0}, t, \mu$ \\
	\textbf{output}: $\mat{X}_{n}$
	\begin{algorithmic}
		\State \textbf{Initialization} Set $F(\mat{X}) = F(\mat{X},t,\mu)$ according to \eqref{SmoothingObj}  and compute
		$\mat{G}_{0}=\grad F(\mat{X}_{0}), \mat{D}_{0}=\mat{G}_{0},\mat{X}_{1}=\mat{X}_{0}.$ 
		\For{$n=1,2,\dots$}
		\State \textbf{1.} Compute $\mat{G}_{n}=\grad F(\mat{X}_{n})$ according to \eqref{grad_F}.
		\State \textbf{2.} \textbf{If} $\norm{\mat{G}_{n}}_F \le \varepsilon_0$ \textbf{then} STOP.
		\State \textbf{3.} Compute $\mat{D}_{n}=\mat{G}_{n}+\nu_{n}\mat{Y}_{n}$  according to \eqref{conjugateDirection} and \eqref{HSrule}. \\
		\qquad\: \textbf{If} $\langle \mat{G}_{n},\mat{D}_{n}\rangle_{\mat{X}_{n}} < 0$ \textbf{then} $\mat{D}_{n}=\mat{G}_{n}$.  
		\State \textbf{4.} Compute $\mat{X}_{n+1}=R_{\mat{X}_{n}}(\alpha_n\mat{D}_{n})$ using retraction in \eqref{retraction}\\
		\qquad\: where $ \alpha_n=\text{ArmijoLineSearch}(\mat{X}_{n},\mat{D}_{n},\mat{X}_{n-1}).$ 
		
		\EndFor
	\end{algorithmic}
\end{algorithm}
\begin{algorithm}
	\caption{$ \alpha =\text{ArmijoLineSearch}(\mat{X},\mat{D},\mat{X}_{0})$}
	\textbf{input}: $ \mat{X},\mat{D}, \mat{X}_{0}(optional)$ \\
	\textbf{output}: $\alpha$
	\begin{algorithmic}
		\State \textbf{If} $\mat{X}_{0}$ is absent \textbf{then} $\alpha = 1/\norm{\mat{D}}_F;$ \\
		\textbf{else} $\alpha = 2 \tfrac{F(\mat{X})-F(\mat{X}_{0})}{\langle \grad F(\mat{X}),\mat{D}\rangle_\mat{X}}.$ 
		\State  \textbf{If} $\alpha \norm{\mat{D}}_F \le 10^{-10}$ \textbf{then} $\alpha = 1/\norm{\mat{D}}_F.$ 
		\While{$F(R_{\mat{X}}(\alpha\mat{D})) - F(\mat{X}) < 10^{-4}\alpha\langle \grad F(\mat{X}),\mat{D}\rangle_\mat{X}$} 
		\State $\alpha = \alpha/2.$ 
		\EndWhile
	\end{algorithmic}
\end{algorithm}

Incorporating the RCG subroutine into the Dinkelbach-type procedure in Algorithm 1, 
the overall algorithm provides monotonically improving approximations to a stationary solution to 
the multicast beamforming problem $(\mathcal{F}_*)$ and guarantees feasibility with a low complexity.

\section{Simulation Results}
In this section, we provide numerical examples to illustrate the performance of the proposed DT-RCG algorithm. 
We consider a multicast scenario consisting of $3$ cells and $10$ single-antenna users per cell.
The intracell and intercell channels are i.i.d. generated from $\mathcal{CN}(0,1)$
and $\mathcal{CN}(0,\epsilon),$ respectively. 
We set $\epsilon=1/4$ which means that the intercell channels undergo $6$ dB
stronger large-scale fading than the intracell channels. 
The noise variance of each user is set as $\sigma_{l,k}^2=1,\forall l,k.$
For the sake of simplicity, we assume that all users in the system have a common SINR target $\Gamma=1$
and the power budget vector for the three BSs is $\vect{P}=[P,P,2P]^T$. 
All results below are averaged over $500$ channel realizations.

\begin{figure}[!t]
	\centering
	\includegraphics[width=0.38\textwidth,height=0.207\textheight]{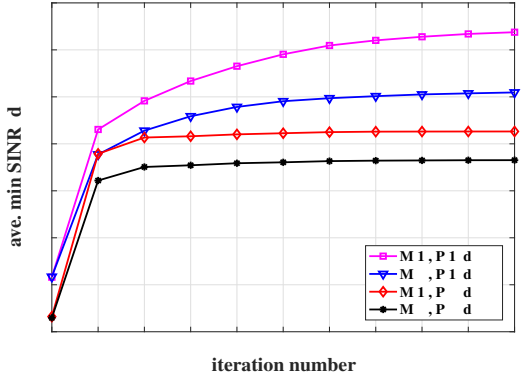}
	\caption{Convergence curve of the DT-RCG algorithm.}
	\label{EPSconvergenceCurve}
\end{figure}

In Fig. \ref{EPSconvergenceCurve}, we illustrate the convergence curve of the DT-RCG algorithm.
The results validate the monotonicity and convergence of the proposed algorithm. 
We can also see that at the first few iterations, the DT-RCG algorithm converges very fast and achieves 
the major part of the limiting value.

\begin{figure}[!t]
	\centering
	\includegraphics[width=0.38\textwidth,height=0.207\textheight]{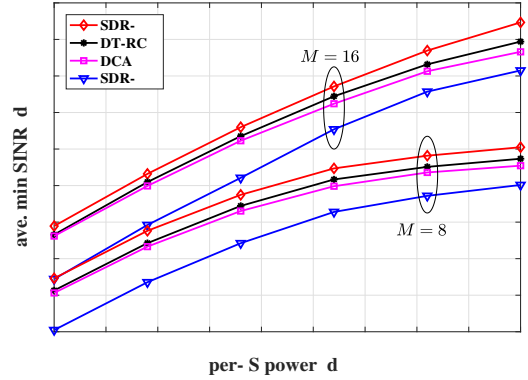}
	\caption{Average minimum SINR versus per-base-station power, $P$.}
	\label{MinSINRvsPower}
\end{figure}

In Fig. \ref{MinSINRvsPower}, we compare the average minimum SINR of the 
DT-RCG algorithm with that of the bisection-based SDR-G scheme \cite{2008TSP-MultiGroup}\cite{2013-coordinated} and
the Difference of Convex-functions Algorithm (DCA) \cite{hsu2016joint}. The SDR upper bound (SDR-UB) of the minimum SINR is also presented.
It can be seen that the average minimum SINR achieved by DT-RCG is close to DCA,  and substantially higher than SDR-G. 
The gap between the achievable minimum SINR for DT-RCG and the SDR upper bound is less than $1$ dB for $M=16$ and less than
$0.6$ dB for $M=8.$ The required number of arithmetic operations per inner iteration for DT-RCG, SDR-G and DCA 
is $\mathcal{O}(L^2MK),\mathcal{O}(L^3M^6+L^2M^2K)$ \cite{2008TSP-MultiGroup} and $\mathcal{O}(L^3M^3+L^4MK)$ \cite{2001-ben-lectures}, respectively. 
Hence, DT-RCG is expected to considerably outperform SDR-G and DCA for a fixed number of outer and inner iterations
\footnote{We found that DT-RCG generally converged within 10 outer iterations and 200 inner iterations for modest accuracy
	and that SDR-G and DCA nearly always converged within 20 outer iterations. It's known that actual iteration complexity 
	of interior point methods for SDP and SOCP is typically about few tens and independent of $L, K$ and $M$.}.

\section{Conclusions}
In this paper, the max-min fair coordinated multicast beamforming under individual BS power constraints
was investigated. The original problem was first recast in a tractable 
parametric programming form. Afterwards, an efficient Riemannian conjugate gradient algorithm was developed 
for each parametric subproblem. The overall algorithm features monotonically improving approximations 
to a stationary solution of the original problem and therefore guarantees convergence.
Numerical results validate the effectiveness of the proposed multicast beamforming algorithm and show 
significant advantages of it over the SDP-based method and the DC-programming-based method in terms of 
both better performance and lower computational complexity.

\ifCLASSOPTIONcaptionsoff
  \newpage
\fi

%








\end{document}